\documentclass[11pt,a4paper]{article}

\usepackage{jheppub}
\usepackage{amsmath,amssymb,amsfonts,amsthm}
\usepackage{amsbsy} 
\usepackage{epsfig}
\usepackage{latexsym}
\input amssym.def
\input amssym.tex

\usepackage{hyperref}

\newcommand{\oarX}[1]{\href{http://arxiv.org/abs/#1}{{\ttfamily #1}}}
\newcommand{\arX}[1]{\href{http://arxiv.org/abs/#1}{{\ttfamily arXiv:#1}}}
\newcommand{\doin}[2]{\href{http://dx.doi.org/#1}{#2}}

\def\be{\begin{equation}}
\def\ee{\end{equation}}
\def\bea{\begin{eqnarray}}
\def\eea{\end{eqnarray}}

\newcommand{\dd}{\mathrm{d}}
\def\im{{\rm i}}

\title{Boundary conditions and Hilbert spaces in no-roll quantum cosmology}
\author{Steffen Gielen}
\affiliation{School of Mathematical and Physical Sciences, University of Sheffield,
Hicks Building, Hounsfield Road, Sheffield S3 7RH, United Kingdom}
\emailAdd{s.c.gielen@sheffield.ac.uk}

\abstract{We construct Hilbert spaces for the minisuperspace quantum cosmology of a closed Universe in the limit of extreme slow-roll inflation, in which the scalar field is approximated as constant. In this setting, the potential energy in the scalar field is an integration constant depending on initial conditions, equivalent to the cosmological constant as it appears in unimodular gravity. If one fixes the value of this integration constant, the Wheeler--DeWitt equation admits two independent solutions, and a natural inner product picks out one of them (essentially Vilenkin's tunnelling wavefunction) with positive norm. The physical Hilbert space is then one-dimensional, in agreement with some recent discussions of closed universes in quantum gravity. However, if the potential energy is left arbitrary, the theory allows for an infinite-dimensional Hilbert space corresponding to energy eigenstates of an effective Hamiltonian. Requiring that this Hamiltonian be represented as a self-adjoint operator leads to a one-parameter family of boundary conditions at the singularity, generalising the DeWitt criterion of a vanishing wavefunction. The boundary condition always leads to a mixture of Hartle--Hawking (no-boundary) and tunnelling wavefunctions, but a particular choice ``almost'' singles out the Hartle--Hawking wavefunction, with exponentially suppressed corrections.}

\begin{document}
\maketitle

\section{Introduction}

Quantum cosmology is an attempt to build a theory of the early Universe based on the principles of quantum mechanics and general relativity. In the absence of a fully workable theory of quantum gravity, the usual starting point is symmetry reduction to minisuperspace: the Universe is taken to be spatially homogeneous and isotropic, perhaps with perturbatively small inhomogeneities. Such an approach can have many goals -- testing techniques for quantising gravity in a simpler setting or investigating conceptual questions -- but one of the most important goals is to provide an initial state for the Universe which can then generate the classical structures we observe. The traditional approach to quantum cosmology was initiated in the early days of canonical quantum gravity \cite{DeWitt} and then advanced to a stage where proposals could be compared \cite{Vilenkin_review, Halliwell_lectures}. In particular, the Hartle--Hawking no-boundary state \cite{HartleHawking} and Vilenkin's tunnelling wavefunction \cite{Tunneling} are two prominent candidates for the initial state of a closed homogeneous and isotropic Universe within slow-roll inflation. 

The explicit form of quantum cosmology wavefunctions is often derived from a gravitational path integral with given boundary conditions and in a semiclassical approximation; one may alternatively view them as particular solutions to the canonical Wheeler--DeWitt equation. Some recent discussions in the field have centred on how exactly the path integral is to be evaluated (starting with \cite{Feldbrugge_PRL, Feldbrugge}). We will instead be concerned with the question of how the ``wavefunction of the Universe'' is to be interpreted: what is the physical Hilbert space of quantum cosmology wavefunctions for a closed Universe with inflation (in the extreme slow-roll limit)? Here, recent developments in the study of gravitational path integrals lead to the claim that the physical Hilbert space of closed universes is one-dimensional \cite{1dHilbert}. Such a claim is rather different from traditional interpretations of quantum cosmology wavefunctions as more similar to the quantum states of a particle, which can take a number of configurations and for which probabilistic statements are meaningful. If the physical Hilbert space is one-dimensional, it does not seem meaningful to talk about transition probabilities between different configurations of such a Universe, as they would all be equal to one \cite{consistent}.

The approach we are taking here is rather conservative, simply based on the quantisation of a homogeneous and isotropic Universe via the Wheeler--DeWitt equation in the extreme ``no-roll'' limit in which the scalar field can be taken to be constant. We observe that in such a situation, the constant potential energy $V(\phi_*)$, with $\phi=\phi_*$ the constant value of the scalar field in the limit, is analogous to the total energy of a mechanical particle system or to the cosmological constant in unimodular gravity, which appears as an integration constant. One may take $V(\phi_*)$ to be a given constant, in which case the classical system has no degrees of freedom, and the space of solutions to the Wheeler--DeWitt equation is two-dimensional. A natural choice of inner product renders one of the solutions, essentially Vilenkin's tunnelling wavefunction, to be of positive norm, while the other one has negative norm and can be discarded. In this situation, our result agrees with the statement that there is only a one-dimensional physical Hilbert space. It is not meaningful to ask for the probability of such a Universe to have a particular size, and the scale factor $a$ is more similar to a time coordinate; as the state ``evolves'' it will eventually reach all large enough values of $a$ with certainty, just as the classical de Sitter solution does. A positive definite choice of inner product allows for both solutions to the Wheeler--DeWitt equation and returns a two-dimensional physical Hilbert space, corresponding to the expanding and contracting classical solutions.

An alternative and perhaps more physical proposal is to include all possible values of $V(\phi_*)$, as one would do in a classical particle system or in unimodular gravity. The challenge is then to build a Hilbert space of all quantum states corresponding to different values of $V(\phi_*)$. Our proposal is to build this Hilbert space on the assumption that the physical Hamiltonian of the minisuperspace cosmology, whose eigenvalues determine the possible values of $V(\phi_*)$, is represented as a self-adjoint operator.\footnote{There is a substantial literature discussing various definitions of self-adjoint Hamiltonians in quantum cosmology, see, e.g., \cite{Fselfadj} for a related paper.} This simple assumption suggests a natural inner product for the wavefunction of the Universe and leads to the imposition of a boundary condition at the singularity $a=0$, somewhat reminiscent of DeWitt's proposal \cite{DeWitt} but coming from a completely different angle: the boundary condition is needed for self-adjointness. We show that such a boundary condition, to be chosen from a one-parameter family of possible boundary conditions, does not single out either the no-boundary state or the tunnelling wavefunction, but always requires a mixture of the two (dependent on the value of $V(\phi_*)$). However, in the physically interesting regime $0<V(\phi_*)\ll 1$ in Planck units one can choose the physical wavefunctions to be ``almost'' given by the no-boundary wavefunction, with a correction that is exponentially small and entirely negligible when the Universe becomes macroscopic. In this sense, we are able to give a new criterion that picks out the no-boundary wavefunction as preferred over its alternatives. We discuss implications for probabilistic interpretations of quantum cosmology, and possible relations to derivations of wavefunctions via the gravitational path integral. Our construction of a theory involving all possible values of $V(\phi_*)$ leads to an infinite-dimensional Hilbert space, inconsistent with the idea that the physical Hilbert space would be one- or two-dimensional.

\section{Minisuperspace quantum cosmology and no-roll limit}

We work in the traditional minisuperspace setting where spacetime $\mathcal{M}=S^3\times\mathbb{R}$ is modelled as a closed ($k=1$) homogeneous and isotropic universe, with metric given by
\be
\dd s^2 = -N^2(t) \,\dd t^2 + a^2(t) \, \dd\Omega^2
\ee
where $\dd\Omega^2$ is the line element on a 3-sphere of unit radius. Since we are interested in inflation, we also include a scalar field $\phi$ with some potential $V(\phi)$, which in this approximation is likewise assumed to be spatially homogeneous. We also choose units in which $8\pi G=1$.

The Einstein--Hilbert action, supplemented by a Gibbons--Hawking--York boundary term \cite{GHY} on the past and future boundary $\partial\mathcal{M}$, together with the standard scalar field action, then reduces to (see, e.g., \cite{Lehners_review,Gielen_review,Feldbrugge})
\bea
S_{{\rm EH}} & = & \int_{\mathcal{M}} \dd^4 x\;\sqrt{-g}\left(\frac{R}{2}-\frac{1}{2}g^{\mu\nu}\partial_\mu\phi\partial_\nu\phi-V(\phi)\right)  + \int_{\partial\mathcal{M}} \dd^3 x\;\sqrt{q}K  \nonumber
\\ & = & 2\pi^2 \int \dd t\;N \left(-3a\frac{\dot{a}^2}{N^2} + 3a + a^3 \frac{\dot\phi^2}{2N^2} - a^3V(\phi)\right)
\eea
where we have set the bare cosmological constant to zero, since it can be absorbed into the potential $V(\phi)$.

In the extreme slow-roll (i.e., a no-roll) limit, one assumes that the kinetic energy in the scalar field can be neglected, and that the scalar sits at an approximately constant value $\phi=\phi_*$, so that the action simplifies to
\be
S_{{\rm EH}} = 2\pi^2 \int \dd t\;N \left(-3a\frac{\dot{a}^2}{N^2} + 3a - a^3V(\phi_*)\right)
\label{simple_action}
\ee
which is now equivalent to the action for pure gravity with cosmological constant $\Lambda=V(\phi_*)$. However, the values of $\phi_*$ and $V(\phi_*)$ are not fixed by any principle. Indeed, one of the main goals of quantum cosmology is to make predictions for possible values of $V(\phi_*)$ and hence possible initial conditions for the Universe. The effective cosmological constant arising from a fixed scalar potential energy in the no-roll limit is akin to the cosmological constant as it appears in unimodular gravity \cite{unimod}, namely as an integration constant fixed by initial conditions, rather than a fundamental constant of Nature. 

The equations of motion derived from (\ref{simple_action}) are the standard Friedmann equations
\bea
\frac{\dot{a}^2}{a^2N^2} + \frac{1}{a^2} &=& \frac{1}{3}V(\phi_*)\,,  
\label{fried1}
\\ \frac{\ddot{a}}{a N^2}-\frac{\dot{a}\dot{N}}{a N^3}   &=&\frac{1}{3} V(\phi_*)
\label{fried2}
\eea
where the second equation is not independent from the first: differentiating (\ref{fried1}) with respect to time  (with the no-roll approximation $\dot\phi_*=0$) yields
\be
\frac{\ddot{a}}{aN^2}=\frac{\dot{a}^2}{a^2N^2}+\frac{\dot{a}\dot{N}}{aN^3}+\frac{1}{a^2}
\label{accel}
\ee
and subtracting (\ref{fried1}) then yields (\ref{fried2}). If we are thinking of $V(\phi_*)$ as determined by initial conditions (and hence arbitrary) rather than a fixed constant, it makes more sense to regard (\ref{accel}) as the fundamental equation of motion. Its integral is then (\ref{fried1}), with the right-hand side a constant analogous to the total energy in standard mechanics.

The connection with unimodular gravity can be made more obvious by starting with (\ref{simple_action}) and imposing the gauge $N=1/a^3$, which corresponds to a fixed time-independent volume element. The action then becomes
\be
S_{{\rm UG}} = 2\pi^2 \int \dd t \left(-3a^4\dot{a}^2 + \frac{3}{a^2} \right)
\label{unimod_action}
\ee
where we drop the term involving $V(\phi_*)$, which is now just an additive constant with no influence on the equations of motion. Variation with respect to $a$ gives
\be
a^4\ddot{a} + 2a^3\dot{a}^2 - \frac{1}{a^3} = 0
\ee
which is the same as (\ref{accel}) for $N=1/a^3$. (\ref{unimod_action}) is a well-defined action for unimodular gravity in minisuperspace, resulting in a dynamical equation whose integration gives (\ref{fried1}) with an undetermined right-hand side, i.e.,
\be
a^4\dot{a}^2 + \frac{1}{a^2} = E = {\rm const.}
\ee
In this theory, we do not impose the trace part of the Einstein equations \cite{ellis_unimod}. This seems to be the right setting for studying all possible no-roll solutions of inflation in minisuperspace; notice that we must have $E>0$ or $V(\phi_*)>0$. 

The Hamiltonian formulations of both theories are similarly closely related. Namely, starting from (\ref{simple_action}) one finds a canonical momentum
\be
p_a = -12\pi^2 \frac{a\dot{a}}{N}
\ee
and a constrained Hamiltonian 
\be
H = N\left(-\frac{p_a^2}{24\pi^2 a}+2\pi^2(-3a+a^3 V(\phi_*))\right)\approx 0\,.
\label{hamconst}
\ee
Similarly, starting from the unimodular gravity action (\ref{unimod_action}) we find a canonical momentum
\be
p_a = -12\pi^2 a^4\dot{a}
\ee
and an {\em unconstrained} Hamiltonian
\be
H_{{\rm UG}} = -\frac{p_a^2}{24\pi^2 a^4} - \frac{6\pi^2}{a^2}
\label{ug_hamilt}
\ee
The expressions for $p_a$ and $H$ in the unimodular setting are formally identical to the ones in general relativity after setting $N=1/a^3$ and dropping the term involving $V(\phi_*)$, taking into account the fact that we no longer have the Hamiltonian constraint, so $H_{{\rm UG}}$ is allowed to take arbitrary values on physical states.

Standard Wheeler--DeWitt quantisation is based on imposing the Hamiltonian constraint (\ref{hamconst}) as a differential equation for a wavefunction $\Psi(a)$. To represent the Hamiltonian as an operator, a choice of operator ordering is needed, which is usually derived from a choice of lapse. If we choose $N=1/a$, a popular choice in the literature \cite{Feldbrugge,Halliwell88}, then
\be
H = -\frac{p_a^2}{24\pi^2 a^2}-2\pi^2(3-V(\phi_*) a^2) = G^{-1}(a) p_a^2 -2\pi^2(3-V(\phi_*) a^2)
\ee
where $G(a)=-24\pi^2 a^2$ can be seen as defining a metric on the one-dimensional minisuperspace parametrised by $a$. The d'Alembert operator with respect to this metric is
\be
\Box_G = -\frac{1}{\sqrt{|G|}}\frac{\dd}{\dd a}\left(\frac{1}{\sqrt{|G|}}\frac{\dd}{\dd a}\right) = -\frac{1}{24\pi^2 a}\frac{\dd}{\dd a}\left(\frac{1}{a}\frac{\dd}{\dd a}\right)
\ee
and by representing the term $G^{-1}(a)p_a^2$ as $-\Box_G$, the Wheeler--DeWitt equation for given $V(\phi_*)$ is
\be
\left[\frac{1}{24\pi^2 a}\frac{\dd}{\dd a}\left(\frac{1}{a}\frac{\dd}{\dd a}\right)-6\pi^2+2\pi^2 V(\phi_*) a^2\right]\Psi(a)=0\,.
\label{wdw}
\ee
This has the famous general solution in terms of Airy functions \cite{Vilenkin_review,Lehners_review},
\be
\Psi(a) = c_1 \, {\rm Ai}\left( z \right) + c_2 \, {\rm Bi}\left( z \right)
\label{psidef}
\ee
with
\be
z = \left(\frac{18\pi^2}{V(\phi_*)}\right)^{2/3}\left(1-a^2 \frac{V(\phi_*)}{3}\right)\,.
\ee
The choice $\Psi(a) \sim {\rm Ai}(z)$ corresponds to the Hartle--Hawking no-boundary wavefunction \cite{HartleHawking} while $\Psi(a) \sim {\rm Ai}(z)+\im\, {\rm Bi}(z)$ corresponds to Vilenkin's tunnelling wavefunction \cite{Tunneling}. The constant of proportionality in these expressions is usually fixed by path-integral arguments and often interpreted probabilistically, as we will discuss further below. In the canonical setting based on the Wheeler--DeWitt equation, $c_1$ and $c_2$ are initially arbitrary constants.

Rewriting the Wheeler--DeWitt equation (\ref{wdw}) as
\be
\left[-\frac{1}{48\pi^4 a^3}\frac{\dd}{\dd a}\left(\frac{1}{a}\frac{\dd}{\dd a}\right)+\frac{3}{a^2}\right]\Psi(a)= V(\phi_*)\Psi(a)\,,
\ee
we can interpret it as an eigenvalue equation for a particular differential operator
\be
\hat{H}_S = -\frac{1}{48\pi^4 a^3}\frac{\dd}{\dd a}\left(\frac{1}{a}\frac{\dd}{\dd a}\right)+\frac{3}{a^2}
\label{HS}
\ee
with eigenvalue $V(\phi_*)$. Viewing $V(\phi_*)$ not as a given fixed constant but a quantity that can take different values for different states then suggests building a Hilbert space out of wavefunctions of the form (\ref{psidef}) for different values of $V(\phi_*)$. 

Again, this is also what one would do after quantising the unconstrained Hamiltonian (\ref{ug_hamilt}) in a particular operator ordering and constructing the time-independent Schr\"odinger equation
\be
\left[-\frac{1}{48\pi^4 a^3}\frac{\dd}{\dd a}\left(\frac{1}{a}\frac{\dd}{\dd a}\right)+\frac{3}{a^2}\right]\Psi_E(a)= -\frac{E}{2\pi^2}\Psi_E(a)
\ee
with $E$ now corresponding to the eigenvalues of $\hat{H}_{{\rm UG}}=-2\pi^2\hat{H}_S $. A corresponding time-dependent Schr\"odinger equation is defined in the standard way as
\be
\left[-\frac{1}{48\pi^4 a^3}\frac{\partial}{\partial a}\left(\frac{1}{a}\frac{\partial}{\partial a}\right)+\frac{3}{a^2}\right]\Psi_E(a,T)= -\frac{\im\hbar}{2\pi^2}\frac{\partial}{\partial T}\Psi(a,T)
\label{new_rohinga}
\ee
with $T$ corresponding to unimodular time (the time coordinate in the gauge $N=1/a^3$). This equation defines unitary evolution in unimodular time if $\hat{H}_S$ is self-adjoint on an appropriate Hilbert space of states.

It is worth noting that (\ref{new_rohinga}) is the not the most natural operator ordering associated to the unimodular gravity Hamiltonian (\ref{ug_hamilt}), which would instead be a symmetric form
\be
\left[-\frac{1}{48\pi^4 a^2}\frac{\partial}{\partial a}\left(\frac{1}{a^2}\frac{\partial}{\partial a}\right)+\frac{3}{a^2}\right]\Psi_E(a,T)= -\frac{\im\hbar}{2\pi^2}\frac{\partial}{\partial T}\Psi(a,T)\,.
\ee
In our model, the procedure of first quantising in an operator ordering suggested by the gauge $N=1/a$ and then rearranging and effectively going to unimodular gauge $N=1/a^3$ yields a different theory from the one obtained directly in unimodular gauge. This factor ordering ambiguity is a standard problem in quantisation, which can be avoided in minisuperspace dimension greater than one by quantising the classical Hamiltonian in a conformally invariant way \cite{Halliwell88,Gielen_review}. Here, since we are dealing with a simple minisuperspace model in which the only degree of freedom is the scale factor $a$, no such preferred ordering can be adopted and one has to live with a dependence of the theory on the choice of gauge. We will keep (\ref{HS}) as the operator ordering appearing in our eigenvalue equation, in order to connect with the literature on the no-boundary and tunnelling wavefunctions. 

The choice of operator ordering that would lead to a Schr\"odinger equation
\be
\left[-\frac{1}{48\pi^4 a^4}\frac{\partial^2}{\partial a^2}+\frac{3}{a^2}\right]\Psi_E(a,T)= -\frac{\im\hbar}{2\pi^2}\frac{\partial}{\partial T}\Psi(a,T)
\ee
was studied from a perspective similar to ours in \cite{dynvacuum}. The authors also discuss the need for boundary conditions to obtain a self-adjoint Hamiltonian. In explicit solutions, they restrict to the spatially flat ($k=0$) case for simplicity, and do not discuss Hartle--Hawking and Vilenkin wavefunctions.

\section{Hilbert spaces of inflationary no-roll wavefunctions}

\subsection{Fixed initial conditions}

We have seen that the space of solutions to the Wheeler--DeWitt equation (\ref{wdw}) for {\em fixed} scalar field potential $V(\phi_*)$ is at most two-dimensional. This is the Hilbert space of a fully constrained theory: classically, there is a single phase space pair of dynamical variables $\{a,p_a\}$ and a single (first-class) constraint. To construct a physical Hilbert space from these solutions, given that they solve a second order differential equation in $a$, an inner product could be constructed from the standard Klein--Gordon conserved current
\be
J[\Psi] = -\im\sqrt{|G|}G^{-1}\left(\bar\Psi(a)\Psi'(a)-\Psi(a)\bar\Psi'(a)\right) = \frac{\im}{\sqrt{24}\pi a}\left(\bar\Psi(a)\Psi'(a)-\Psi(a)\bar\Psi'(a)\right)
\label{KGnorm}
\ee
This should be independent of $a$, consistent with a view of the Wheeler--DeWitt equation as a second-order evolution equation in ``time'' $a$. Indeed, evaluating $J$ explicitly for a wavefunction of the form (\ref{psidef}), one finds
\be
J[\Psi] = -\im \left(\frac{\sqrt{2}V(\phi_*)}{\sqrt{3}\pi^2}\right)^{1/3} \left(\bar{c}_1 c_2-c_1\bar{c}_2 \right)\,,
\ee
and so the inner product is $a$-independent but not positive definite, as is expected. 

In this inner product, a normalised state with positive norm, $J[\Psi]=1$, is given by
\be
\Psi(a) = \left(\frac{\sqrt{3}\pi^2}{\sqrt{2}V(\phi_*)}\right)^{1/6}\left(\frac{1}{\sqrt{2}}{\rm Ai}(z) + \frac{\im}{\sqrt{2}}{\rm Bi}(z)\right)
\ee
whereas a negative-norm state with $J[\Psi]=-1$ is given by the complex conjugate,
\be
\Psi(a) = \left(\frac{\sqrt{3}\pi^2}{\sqrt{2}V(\phi_*)}\right)^{1/6}\left(\frac{1}{\sqrt{2}}{\rm Ai}(z) - \frac{\im}{\sqrt{2}}{\rm Bi}(z)\right)\,.
\ee
The two states are orthogonal. The positive-norm state is (up to normalisation) given by Vilenkin's tunnelling wavefunction, which one could hence declare to be the only allowed physical state of the theory. The theory is unitary in a somewhat trivial sense, given that there is only one state and no transitions are possible.

In this simple minisuperspace setting, such a result agrees with recent statements in the literature that the physical Hilbert space of closed universes is one-dimensional (see, e.g., \cite{1dHilbert}), which are often derived in a statistical interpretation of the gravitational path integral. In our setting, we find a one-dimensional Hilbert space in the simplest minisuperspace model with fixed initial conditions for the scalar field, but we would not expect such a result to hold more generally once more degrees of freedom are added. The minisuperspace model at fixed $V(\phi_*)$ has zero degrees of freedom -- the unique solution is given by de Sitter space in the given closed slicing. Gravity beyond minisuperspace has additional local degrees of freedom. In our more conventional approach, we make no claim to replicate the argument for one-dimensional gravitational Hilbert spaces more generally.

The perspective advocated here does however strongly differ from the traditional perspective in minisuperspace quantum cosmology which one sees $\Psi(a)$ as more similar to a wavefunction of an unconstrained particle in one dimension, which has one degree of freedom and where probabilistic statements about different values of $a$ might be meaningful. If we view $a$ as analogous to time and the physical Hilbert space as one-dimensional, the state simply {\em is}; it corresponds to the one allowed classical configuration (de Sitter space), which is observed with probability one. The classical solution will eventually reach arbitrarily large values of $a$ -- all sufficiently large values of $a$ are related by a gauge transformation generated by the Hamiltonian constraint -- and so the probability for reaching some large enough value of $a$ should again always be equal to one. Again, this observation is also compatible with the more general result derived in a path-integral setting in \cite{consistent}.

There is another possible point of view, in which we adopt a different inner product and take the physical Hilbert space to be two-dimensional. We could simply declare both the tunnelling wavefunction and its complex conjugate to be states of positive norm. Such an inner product corresponds to an inner product constructed from group averaging \cite{groupav} in standard Dirac (constraint) quantisation \cite{Diracconst}, where one inserts a projector $\delta(\hat{C})$, where $\hat{C}$ is a constraint operator, into a kinematical inner product. The resulting physical inner product is positive definite (see, e.g., \cite{HoehnSwitch} for an explicit example of such a construction). In such a Hilbert space, the distinction between the two states, loosely associated with an expanding and a contracting de Sitter universe, is meaningful, and the superposition of both that leads to the no-boundary state would also be a well-defined state of positive norm, unlike for the Klein--Gordon-like inner product (\ref{KGnorm}) in which a real wavefunction (such as the no-boundary wavefunction) is associated with zero norm. In a two-dimensional physical Hilbert space, the state of (the gauge equivalence class of) some large $a$ is not necessarily equivalent to a no-boundary state.

\subsection{Arbitrary initial conditions and self-adjoint Hamiltonian}

The situation is different if we try to build a Hilbert space of no-roll wavefunctions for all possible values of $V(\phi_*)$, as would seem more physically reasonable given that different initial conditions for $\phi_*$ are possible. The previous discussion has shown that these are analogous to energy eigenstates of a Hamiltonian $\hat{H}_S$ defined in (\ref{HS}), which acts on wavefunctions $\Psi(a)$. In this setting, there is no Hamiltonian constraint. To define a Hilbert space, we then need to again fix an inner product, and it seems natural to require that $\hat{H}_S$ be represented as a self-adjoint operator. 

A good candidate for an inner product with respect to which this is possible is
\be
\langle\Psi|\Phi\rangle = \int_0^\infty \dd a \;a^3\,\bar\Psi(a)\Phi(a)\,.
\label{innerprod}
\ee
In this inner product, 
\be
\langle\Psi|\hat{H}_S\Phi\rangle = \langle\hat{H}_S\Psi|\Phi\rangle + \frac{1}{48\pi^4}\left[\frac{1}{a}\bar\Psi'(a)\Phi(a) - \frac{1}{a}\bar\Psi(a)\Phi'(a)\right]_0^\infty\,.
\label{symmetry}
\ee
For the operator to be at least symmetric, one needs to ensure that the boundary terms vanish. This is a nontrivial requirement on the allowed wavefunctions, as one can see from verifying the small $a$ behaviour of the general solution (\ref{psidef}). Assuming the physically most interesting case of a positive inflationary potential that is small in Planck units, $0<V(\phi_*)\ll 1$, we have 
\begin{align}
\Psi(a)& \stackrel{a\rightarrow 0}{\sim}\;\; e^{-\frac{12\pi^2}{V(\phi_*)}}\frac{1}{2}\left(\frac{V(\phi_*)}{18\pi^5}\right)^{1/6}(c_1-\im c_2) + e^{\frac{12\pi^2}{V(\phi_*)}}c_2\left(\frac{V(\phi_*)}{18\pi^5}\right)^{1/6}\nonumber
\\ & \qquad + \left[e^{-\frac{12\pi^2}{V(\phi_*)}}3\pi^2 \left(\frac{V(\phi_*)}{18\pi^5}\right)^{1/6}(c_1-\im c_2)  - e^{\frac{12\pi^2}{V(\phi_*)}}6\pi^2 c_2\left(\frac{V(\phi_*)}{18\pi^5}\right)^{1/6} \right]a^2
\label{smalla}
\end{align}
up to higher orders in $a$. We see that both $\Psi(a)$ and $\frac{1}{a}\Psi'(a)$ near the origin can be chosen freely, and the boundary term at $a=0$ is in general nonzero. (This conclusion also holds for larger Planckian or super-Planckian values for $V(\phi_*)$, but we will be less interested in these cases -- they would presumably signal a regime where a simple quantum cosmology based on the Einstein--Hilbert action no longer holds.) 

At large $a$, we have
\begin{align}
\Psi(a)&  \sim \frac{c_1}{(12 V(\phi_*))^{1/12} \pi^{5/6}}\sqrt{\frac{1}{a}}\sin\left(\frac{4\pi^2}{\sqrt{3}V(\phi_*)} \left(a^2 V(\phi_*)-3\right)^{3/2}+\frac{\pi}{4} \right) \nonumber
\\ & \quad + \frac{c_2}{(12 V(\phi_*))^{1/12} \pi^{5/6}}\sqrt{\frac{1}{a}}\cos\left(\frac{4\pi^2}{\sqrt{3}V(\phi_*)} \left(a^2 V(\phi_*)-3\right)^{3/2}+\frac{\pi}{4}\right)
\label{largea}
\end{align}
and hence the boundary term from $a\rightarrow\infty$ has the standard oscillatory plane wave behaviour for free particle eigenstates, and is not expected to make a contribution.

To make the boundary term at $a=0$ vanish, we can impose a Robin-type boundary condition
\be
\Psi(0) = \gamma \lim_{a\rightarrow 0}\frac{1}{a}\Psi'(a)\,,\quad \gamma\in\mathbb{R}\cup \{\infty\}\,.
\label{Robinbound}
\ee
$\gamma$ is a free parameter, characterising the possible self-adjoint extensions of the Hamiltonian $\hat{H}_S$, which translate to possible boundary conditions at $a=0$. The particular choice of $\gamma=0$ is the requirement that the wavefunction vanish at $a=0$, a proposal going back to DeWitt as a possible criterion for resolution of a classical singularity \cite{DeWitt}. Here, we do not interpret the wavefunction physically or make reference to singularity resolution, but simply impose self-adjointness on $\hat{H}_S$.

The same conclusion regarding possible boundary conditions can be reached using the standard method of deficiency indices \cite{deficiency}. First of all, one can make $\hat{H}_S$ into a symmetric operator by imposing the separate boundary conditions $\Psi(0)=\lim_{a\rightarrow 0}\frac{1}{a}\Psi'(a)=0$. In this case, the adjoint satisfies no further restrictions and acts on the original space of normalisable wavefunctions. $\hat{H}_S$ defined in this way is not self-adjoint, as the adjoint has a larger domain than $\hat{H}_S$ itself. To find the self-adjoint extensions, one needs to find normalisable eigenfunctions of this adjoint operator with eigenvalues $\pm \im$, and count the possible unitary maps between the respective eigenspaces.

Eigenfunctions with eigenvalues $\pm\im$ are given by
\be
\Psi_\pm(a)=c_1 \,{\rm Ai}(z_\pm) + c_2 \,{\rm Bi}(z_\pm)\,, \quad z_\pm = \left(\mp 18\im\pi^2\right)^{2/3}\left(1\mp \frac{\im a^2}{3}\right)\,.
\ee
At large $a$, these behave as (dropping subleading terms in the exponent)
\begin{align}
\Psi_+(a) & \sim N_1\frac{c_1+\im\,c_2}{\sqrt{a}}\,e^{(-2+2\im)\sqrt{\frac{2}{3}}\pi^2 a^3} +  N_2\frac{c_1-\im\,c_2}{\sqrt{a}}\,e^{(2-2\im)\sqrt{\frac{2}{3}}\pi^2 a^3}\,,\nonumber
\\ \Psi_-(a) & \sim N_3\frac{c_1-\im\,c_2}{\sqrt{a}}\,e^{-(2+2\im)\sqrt{\frac{2}{3}}\pi^2 a^3} +  N_4\frac{c_1+\im\,c_2}{\sqrt{a}}\,e^{(2+2\im)\sqrt{\frac{2}{3}}\pi^2 a^3}\,,
\end{align}
where $N_i$ are numerical $O(1)$ factors. The Airy functions do not have singularities, and hence we see that the subspaces of normalisable solutions for $\Psi_+$ and $\Psi_-$ are each one-dimensional. The unitary maps between them amount to multiplication by a phase $e^{\im\alpha}$, and this phase is equivalent to the parameter $\gamma$ appearing in the boundary condition (\ref{Robinbound}).

The need to apply boundary conditions to define a self-adjoint extension of the Hamiltonian often arises when the classical evolution has singularities. In quantum cosmology or for quantum black holes, a typical example is when a curvature singularity is encountered in finite time as measured by a particular clock; then usually the Hamiltonian defining evolution in this clock is not symmetric, and boundary conditions are required to characterise the self-adjoint extensions. These boundary conditions, in turn, extend the evolution past the classical singularity \cite{singres}. The case here is different; the classical solutions are still just de Sitter space in closed slicing, which has no coordinate or curvature singularities. The self-adjoint extensions found here apply to the particular factor ordering adopted in (\ref{HS}). This factor ordering has other advantages, notably exact solution in terms of Airy functions rather than more complicated Heun functions, and here it leads to an interesting new conclusion: physical wavefunctions are not arbitrary linear combinations of Airy functions but must obey a boundary condition (\ref{Robinbound}).

\

One might hope that this construction of a self-adjoint Hamiltonian can single out particular choices such as the Hartle--Hawking no-boundary wavefunction or the tunnelling wavefunction, but the situation is more complicated. This is because the relation between the Airy functions and their derivatives at $a=0$ is sensitive to the value of $V(\phi_*)$:
\begin{align}
\lim_{a\rightarrow 0}\frac{\frac{1}{a}\frac{\dd}{\dd a}{\rm Ai}(z)}{{\rm Ai}(z)} & = 12\pi^2 \frac{K_{\frac{2}{3}}\left(\frac{12\pi^2}{V(\phi_*)}\right)}{K_{\frac{1}{3}}\left(\frac{12\pi^2}{V(\phi_*)}\right)} = 12\pi^2 + \frac{V(\phi_*)}{6} - \frac{5V(\phi_*)^2}{864\pi^2} + O\left(V(\phi_*)^3\right)\,,\nonumber
\\ \lim_{a\rightarrow 0}\frac{\frac{1}{a}\frac{\dd}{\dd a}{\rm Bi}(z)}{{\rm Bi}(z)} & = - 12\pi^2 + \frac{V(\phi_*)}{6} + \frac{5V(\phi_*)^2}{864\pi^2} + O\left(V(\phi_*)^3\right)\,,\nonumber
\end{align}
where $K_\nu(z)$ are modified Bessel functions. Since these relations depend on $V(\phi_*)$, the translation of a choice of $\gamma$ parameter in (\ref{Robinbound}) into the allowed linear combination of Airy functions is likewise a function of $V(\phi_*)$. 

In the physically interesting regime $V(\phi_*)\ll 1$, it is however possible to ``almost'' only work with no-boundary type wavefunctions built from the Airy function of the first kind ${\rm Ai}(z)$. This is because, as we can see in (\ref{smalla}), these functions are exponentially small near $a=0$. They also satisfy, to a very good approximation,
\be
\lim_{a\rightarrow 0}{\rm Ai}(z) \approx \frac{1}{12\pi^2} \lim_{a\rightarrow 0}\frac{1}{a}\frac{\dd}{\dd a}{\rm Ai}(z)\,.
\ee
Hence, for the Airy function of the first kind the boundary terms at $a=0$ in (\ref{symmetry}) are very small, even if not exactly zero. 

To see how this works in practice, let us define
\be
\Psi_{{\rm HH},\lambda}(a) = {\rm Ai}\left(z|_{V(\phi_*)=\lambda} \right) \,.
\ee
Recall that these wavefunctions are real-valued. We then find
\begin{align}
& (\lambda - \lambda')\int_0^\infty \dd a\;a^3\;\Psi_{{\rm HH},\lambda}(a)\Psi_{{\rm HH},\lambda'}(a)\nonumber
\\ = & \int_0^\infty \dd a\;a^3\;\left(\left(\hat{H}_S\Psi_{{\rm HH},\lambda}(a)\right)\Psi_{{\rm HH},\lambda'}(a) - \Psi_{{\rm HH},\lambda}(a)\left(\hat{H}_S\Psi_{{\rm HH},\lambda'}(a)\right)\right) \nonumber
\\ = & \frac{1}{48\pi^4}\left[ -\frac{1}{a}\frac{\dd\Psi_{{\rm HH},\lambda}(a)}{\dd a}\Psi_{{\rm HH},\lambda'}(a)  + \frac{1}{a}\Psi_{{\rm HH},\lambda}(a)\frac{\dd\Psi_{{\rm HH},\lambda'}(a)}{\dd a}\right]_0^\infty \nonumber
\\ = & \lim_{x\rightarrow \infty} (\sqrt{\lambda}+\sqrt{\lambda'})\frac{\sin(\frac{4\pi^2(\sqrt{\lambda}-\sqrt{\lambda'})}{\sqrt{3}}x)}{8(18\pi^{11})^{1/3}(\lambda\lambda')^{1/12}}  - \lim_{x\rightarrow \infty} (\sqrt{\lambda}-\sqrt{\lambda'})\frac{\sin(\frac{4\pi^2(\sqrt{\lambda}+\sqrt{\lambda'})}{\sqrt{3}} x)}{8(18\pi^{11})^{1/3}(\lambda\lambda')^{1/12}}\nonumber
\\ &- \frac{1}{48\pi^4}\lim_{a\rightarrow 0}\left( -\frac{1}{a}\frac{\dd\Psi_{{\rm HH},\lambda}(a)}{\dd a}\Psi_{{\rm HH},\lambda'}(a)  + \frac{1}{a}\Psi_{{\rm HH},\lambda}(a)\frac{\dd\Psi_{{\rm HH},\lambda'}(a)}{\dd a}\right) \,;
\end{align}
using the standard relation
\be
\lim_{x\rightarrow\infty}\frac{\sin(kx)}{k} = \pi\delta(k)\,,
\ee
we obtain an approximate orthogonality relation
\be
\int_0^\infty \dd a\;a^3\;\Psi_{{\rm HH},\lambda}(a)\Psi_{{\rm HH},\lambda'}(a) \approx  \frac{\delta\left(\sqrt{\lambda}-\sqrt{\lambda'}\right)}{8(18\pi^{8})^{1/3}\lambda^{1/6}}  \,.
\ee
In this approximate relation we drop the boundary terms coming from $a=0$, which become
\begin{align}
& - \frac{1}{48\pi^4(\lambda-\lambda')}\lim_{a\rightarrow 0}\left( -\frac{1}{a}\frac{\dd\Psi_{{\rm HH},\lambda}(a)}{\dd a}\Psi_{{\rm HH},\lambda'}(a)  + \frac{1}{a}\Psi_{{\rm HH},\lambda}(a)\frac{\dd\Psi_{{\rm HH},\lambda'}(a)}{\dd a}\right)\nonumber
\\ \approx & \; e^{-\frac{12\pi^2}{\lambda}}e^{-\frac{12\pi^2}{\lambda'}}\frac{(\lambda\lambda')^{1/6}}{1152(18\pi^{17})^{1/3}}\,.
\end{align}
For $0<\lambda,\lambda'\ll 1$, these terms are strongly exponentially suppressed. 

A theory with a properly (not just approximately) self-adjoint Hamiltonian would rely on eigenstates that are not of the simple Hartle--Hawking form, but contain a very small correction using the Airy function of the second kind ${\rm Bi}(z)$. Such states can then be normalised to satisfy the relation
\be
\langle\lambda|\lambda'\rangle = \delta(\sqrt{\lambda}-\sqrt{\lambda'})\,.
\ee
In particular, a state of fixed $\lambda$, such as the usual definition of the no-boundary wavefunction, is not normalisable. One could of course define a normalisable state from superposition of a narrow band of $\lambda$ values.

At large $a$, in the regime of physical interest the correction terms including ${\rm Bi}(z)$ will be extremely small, given that the Airy function of the second kind is very large at small $a$ but then has the same fall-off and oscillatory behaviour as the Airy function of the first kind (compare (\ref{smalla}) and (\ref{largea})). Hence, at large $a$ we expect
\be
\Psi_\lambda(a) \sim 3^{1/4}\,\sqrt{\frac{8\pi}{a}}\sin\left(\frac{4\pi^2}{\sqrt{3}\lambda} \left(a^2 \lambda - 3\right)^{3/2}+\frac{\pi}{4}\right)
\ee
for appropriately normalised allowed $\lambda$ eigenstates.

\

It remains to be seen whether there any physical states with $\lambda<0$. In this case, the general solution (\ref{psidef}) to the Wheeler--DeWitt equation becomes
\be
\Psi(a) = c_1 \, {\rm Ai}\left( z \right) + c_2 \, {\rm Bi}\left( z \right)
\ee
with
\be
z = \left(-\frac{18\pi^2}{|\lambda|}\right)^{2/3}\left(1+a^2 \frac{|\lambda|}{3}\right)\,.
\ee
At large $a$, the wavefunction now asymptotically behaves as
\begin{align}
\Psi(a) & \sim \frac{e^{-4\pi^2\sqrt{\frac{|\lambda|}{3}}a^3}}{\sqrt{a}} \frac{(1+\im)}{(-768\pi^{10}|\lambda|)^{1/12}} c_2 \nonumber
\\ & \qquad + \frac{e^{+4\pi^2\sqrt{\frac{|\lambda|}{3}}a^3}}{\sqrt{a}}\frac{6-2\im\sqrt{3}}{8(8748\pi^{10}|\lambda|)^{1/12}}(c_1+\im c_2)\,.
\end{align}
Hence, we can only obtain a normalisable state if $c_2=\im c_1$.  

Comparing the one allowed wavefunction for given $|\lambda|$ to its derivative at $a=0$, one then finds  
\be
\lim_{a\rightarrow 0}\frac{\frac{1}{a}\frac{\dd}{\dd a}({\rm Ai}(z)+\im\,{\rm Bi}(z))}{{\rm Ai}(z)+\im\,{\rm Bi}(z)}  = -12\pi^2\frac{K_{\frac{2}{3}}\left(\frac{12\pi^2}{|\lambda|}\right)}{K_{\frac{1}{3}}\left(\frac{12\pi^2}{|\lambda|}\right)} = -12\pi^2 - \frac{|\lambda|}{6} + \frac{5|\lambda|^2}{864\pi^2} + O\left(|\lambda|^3\right)
\ee
which depends on $\lambda$ in a way that is similar to the positive $\lambda$ case. Since we no longer have two independent coefficients $c_1$ and $c_2$ to be adjusted, whether there are any negative $\lambda$ states at all depends on the choice of boundary condition (\ref{Robinbound}); even for a suitable choice of boundary condition one would expect only one such state. For the choice $\gamma=\frac{1}{12\pi^2}$ that leads to solutions close to the Hartle--Hawking no-boundary wavefunction, there are no negative $\lambda$ states, in agreement with the classical theory that has no solutions for $V(\phi_*)<0$.

\section{Probabilistic statements and path-integral definitions}

We have considered two possible definitions of a Hilbert space for inflationary no-roll wavefunctions within the truncation to minisuperspace. In the first case, where we consider fixed initial conditions and hence a fixed value $V(\phi_*)$, the Hilbert space is either one- or two-dimensional. In the one-dimensional case, we reproduce the results argued for in much more generality (and using different and more general arguments) across the recent literature \cite{1dHilbert, consistent}. There is then just one state in the theory, and it just {\em is}. There are no meaningful probabilistic statements, as anything realised by the one state occurs with probability one. In particular, large $a$ is reached with probability one, as it is in the classical de Sitter space solution represented by this state.

As we have already mentioned above, this is a rather different perspective from the traditional one in quantum cosmology (see, e.g., \cite{Lehners_review}), where one views both the no-boundary wavefunction and alternatives such as the tunnelling wavefunction as arising from a path integral evaluated in the saddle-point approximation. In the case of the no-boundary state, the chosen saddle point is a solution that connects an initial Euclidean regime to a Lorentzian solution representing de Sitter space. The action appearing in the exponent is complex, and one ends up with the result
\be
\Psi_{{\rm nb-PI}}(a) \sim \exp\left(\frac{12\pi^2}{V(\phi_*)}\right)e^{\pm \im \left( \frac{4\pi^2}{\sqrt{3}V(\phi_*)} \left(a^2 V(\phi_*)-3\right)^{3/2} \right)}
\label{nobound_wave}
\ee
where the sign in the exponent depends on which saddles are included (in the original proposal, two saddles with complex conjugate action are included so that the resulting wavefunction is real \cite{Lehners_review, Feldbrugge}). The important result here is the huge exponential enhancement factor, which diverges as $V(\phi_*)\rightarrow 0$. Traditionally, this result was directly interpreted probabilistically as stating that $|\Psi_{{\rm nb-PI}}(a)|^2$ is proportional to the probability of obtaining a particular value $V(\phi_*)$, without considering a possibly $V(\phi_*)$-dependent normalisation factor (see \cite{consistent} for a recent summary and critique of this traditional viewpoint).

\

In our setting, we have shown how to construct a Hilbert space of states for different values of $V(\phi_*)$. The construction depends on the boundary condition at $a=0$ used to achieve self-adjointness of the operator $\hat{H}_S$, but we saw that there is at least one simple choice which means the allowed states are ``almost'' given by the Airy functions appearing in the no-boundary state. Regardless of which boundary condition is chosen, at large $a$ the resulting normalised wavefunctions show simple oscillating behaviour modulated by $\sqrt{1/a}$, where the extra factor is needed because of the inner product (\ref{innerprod}). (The choice of boundary condition typically results in an additional phase shift in the allowed solutions.)

A general state in such a theory, only containing $V(\phi_*)>0$ states, is then given by
\be
|\Psi\rangle = \int_0^\infty {\rm d}\lambda \;A(\lambda)|\lambda\rangle\,, \quad 2\int_0^\infty {\rm d}\lambda \;\sqrt{\lambda}|A(\lambda)|^2 = 1\,. 
\ee
In order to mimic the no-boundary wavefunctions usually derived from a path integral, we could propose
\be
A(\lambda) \sim \exp\left(\frac{12\pi^2}{\lambda}\right)
\ee
which will evidently not lead to a normalisable state if we extend $\lambda$ from zero to infinity: we see that there is both a very bad IR divergence at small $\lambda$, and a milder UV divergence at large $\lambda$. The latter is perhaps not a major problem, given that we do not expect quantum cosmology to be a reliable approximation at arbitrarily high energies, so a UV cutoff needs to be introduced. Since the probability distribution induced by $A(\lambda)$ is very strongly weighted towards small $\lambda$, the exact choice of UV cutoff will not be very important. The theory is then however extremely sensitive to the choice of IR cutoff, and there seem to be no obvious physical arguments for fixing the cutoff. In this sense, no-boundary wavefunctions derived from path integral arguments do not lead to a predictive theory, but only say that $V(\phi_*)$ is as close to zero as allowed, echoing old arguments \cite{Hawking_Lambda}. 

This conclusion is similar to the one reached by interpreting $|\Psi_{{\rm nb-PI}}(a)|^2$ directly as a probability for observing a universe with a certain value of $V(\phi_*)$, as seems to be done implicitly in much of the literature (see again, e.g., \cite{consistent} for a review). Also in that case, one would obtain a probability distribution that diverges very badly in the limit $V(\phi_*)\rightarrow 0$ and, in order to be well-defined, requires an IR cutoff of unclear physical origin. In this sense, our more rigorous construction of a Hilbert space of no-roll inflationary wavefunctions does not lead to new physical insights, given that the natural basis of $V(\phi_*)$ eigenstates consists of states that behave effectively as plane waves at large $a$.

\

A very different conclusion is reached if we focus on the path-integral definition of the tunnelling wavefunction, where a different saddle point is chosen and one finds instead
 \be
\Psi_{{\rm tun-PI}}(a) \sim \exp\left(-\frac{12\pi^2}{V(\phi_*)}\right)e^{\pm \im \left( \frac{4\pi^2}{\sqrt{3}V(\phi_*)} \left(a^2 V(\phi_*)-3\right)^{3/2} \right)}\,.
\ee
This saddle point gives the leading order approximation to the full path integral if the path integral is originally defined as Lorentzian and one picks the leading saddle point using Picard--Lefschetz theory, as emphasised in \cite{Feldbrugge,Feldbrugge_PRL}. This choice suggests we would now have to set
\be
A(\lambda) \sim \exp\left(-\frac{12\pi^2}{\lambda}\right)\,.
\ee
This distribution is now strongly peaked near the UV cutoff at $\lambda\simeq 1$, making the theory very sensitive to the precise value of the cutoff. The theory would predict that $V(\phi_*)$ is as large as allowed, obviously inconsistent with our experience of the Universe. Again, the Hilbert space we have constructed here does not really shed new light on how initial conditions for $V(\phi_*)$ can be determined via path integrals in quantum cosmology.

\section{Conclusions}

We have proposed new constructions of a physical Hilbert space in the simplest traditional setting for quantum cosmology, where one considers a scalar field in a closed homogeneous and isotropic Universe, and the scalar field is approximated to be constant in time (no-roll limit). The guiding principle we used was unitarity, or the construction of an inner product that is independent of time, but what ``time'' means is always a subtle issue in the timeless setting of the Wheeler--DeWitt equation \cite{Kuchar}. We looked at two distinct cases.

In the first case, the potential energy is regarded as fixed, in which case one finds the standard Hamiltonian constraint and no physical degrees of freedom -- the unique solution is de Sitter space with the given effective cosmological constant. If a standard Klein--Gordon inner product is chosen, only one of these states has a positive norm, but a positive definite choice of inner product is also possible. The physical Hilbert space is one- or two-dimensional. In this approach, ``time'' is the only dynamical variable in the theory, the scale factor $a$. Only in this case, our results can agree with recent statements in the literature that the physical Hilbert space of closed universes is one-dimensional \cite{1dHilbert}.

In the second case, the potential energy is not regarded as fixed but as an integration constant, which frees up a dynamical degree of freedom; there is a nontrivial choice of initial conditions for the scalar field, which determines the possible solutions. This is arguably the physically more relevant setting for quantum cosmology in inflation, where we do not regard initial conditions as fixed once and for all but are rather interested in a quantum theory that could predict initial conditions. In this setting, we have an effective Hamiltonian, which can be seen as the Hamiltonian of unimodular gravity in minisuperspace. This Hamiltonian is unconstrained, and its eigenstates span an infinite-dimensional Hilbert space. We propose to use self-adjointness of the Hamiltonian as a physical criterion for determining an inner product and a possible one-parameter family of boundary conditions at the singularity. This one-parameter family of boundary conditions implies a new condition relating the wavefunction and its derivative at the singularity, which generalises the often used DeWitt requirement of a wavefunction that vanishes at $a=0$. The origin of this boundary condition is not in a requirement of singularity resolution, but rather in the requirement of a self-adjoint Hamiltonian or unitarity in the (unimodular) time conjugate to the scalar field potential energy. One possible choice of boundary condition ``almost'' picks out the Hartle--Hawking no-boundary state, with exponentially small corrections.

Our results provide a novel angle on the construction of physical Hilbert spaces in quantum cosmology, based only on basic principles of quantum mechanics and without reference to any specific completion within quantum gravity. Our constructions would become significantly more complicated if one goes beyond minisuperspace, although the presence of an effective cosmological constant arising from a scalar field would more generally allow for an embedding into unimodular gravity or, in other words, for a definition of unitarity in terms of the time coordinate conjugate to this cosmological constant \cite{singres, WaldUnruh}. Such an approach would always allow for an effective superposition of different values of $\Lambda$, providing an alternative to constructions such as \cite{dsHilbert} which are based on a fixed value of $\Lambda$. If unitarity is a useful guiding principle in quantum cosmology as it is in quantum mechanics, one would expect boundary conditions at the singularity to arise for a much more general class of models. It is also worth pointing out that the discussion of boundary conditions and unitarity for the Hartle--Hawking and Vilenkin wavefunctions is different if one works with the connection rather than the metric as the fundamental variable \cite{Magueijo:2020ugp}. 

\

{\em Acknowledgements} -- The research leading to these results was partially funded by the Royal Society through the University Research Fellowship Renewal URF$\backslash$R$\backslash$221005.

\end{document}